\def\CHIClassOptions{sigconf,nonacm}
\ifdefined\CHIClassOptions\else
  \def\CHIClassOptions{sigconf,nonacm}
\fi
\edef\CHILoadClass{\noexpand\documentclass[\CHIClassOptions]{acmart}}
\CHILoadClass
\newcommand{\IncludePDFWithOwnGroup}[2][]{\includegraphics[#1]{#2}}

\newcommand{\DiaryAlt}{A vertical timeline presents 43 panic-like symptom episodes from July 2023 to June 2026. Time runs down the page. Every episode has a unique label, E01 through E43, printed inside its main symbol. The left column contains 16 green circles for episodes managed without BZD or emergency care. The middle column contains 25 yellow pill shapes for episodes involving BZD under the existing care plan but no emergency care. The right column contains two star shapes for emergency-care episodes. Drawings above the symbols indicate settings: for example, a house indicates home, a briefcase work, a train rail travel, and an aeroplane air travel. Badges below each symbol record ChatGPT use, marked G, and contact with another person, marked H. Slashes indicate no use or contact; one episode has no H because contact is unclear. ChatGPT use is recorded in 31 episodes and human contact in 16. Dashed connectors connect displaced symbols to calendar dates. Emergency-care events are marked on 15 September 2024 and 17 June 2025. A right-hand rail shows routine SSRI maintenance paused from late 2024 and resumed on 18 June 2025. Quantitative follow-up ends in May 2026; the episode diary continues into June. As-needed BZD remains in the care plan throughout. BZD means benzodiazepine; SSRI means selective serotonin reuptake inhibitor.}
\newcommand{\RoleMatrixAlt}{A five-by-five matrix reads from the role before a switch in each row to the role after it in each column. The row and column order is E, Explainer; R, Reassurance provider; A, Action guide; S, Safety checker; and I, Interrupter of compulsive self-checking. Diagonal dashes exclude unchanged roles. From E: 31 to R, 19 to A, 21 to S, and 10 to I. From R: 18 to E, 20 to A, 3 to S, and 7 to I. From A: 12 to E, 0 to R, 2 to S, and 8 to I. From S: 15 to E, 27 to R, 6 to A, and 2 to I. From I: 5 to E, 5 to R, 11 to A, and 0 to S. These counts total 222 switches. Darker cells indicate larger counts; coral outlines compare S to R, 27, with R to S, 3. A separate key lists primary-role reply totals: E 207, R 114, A 63, S 155, and I 97, totalling 636 replies. Reply totals count role assignments, not switches.}
\newcommand{\ConversationAlt}{Three side-by-side panels show condensed English translations of selected conversation excerpts. In E15, the participant reports that the suggested breathing step cannot be used, and later replies place greater weight on warning signs and urgent care. In E24, the participant reports repeated measurement; later replies propose stopping and a future rule for checking. In E06, explanation, reassurance, and safety questions remain intertwined as the feared condition changes. Read each panel from top to bottom, then move from the left panel, E15, to the centre, E24, and the right, E06. Every turn names its speaker; bold role labels and arrows identify the two role changes, so colour is not needed to follow them. E15 ends with an urgent-care proposal; E24 ends with a stopping rule, without a recorded outcome. E06 ends with overlapping roles and explicitly labels the recorded paper-bag rebreathing advice as unsafe. The complete displayed translations remain selectable text within the figure.}
\newcommand{\ScaleAlt}{A two-panel longitudinal figure based on psychological scale records generated through the first author's participation in a three-year national gut--brain-axis follow-up programme. The upper panel, A, overlays seven dated SCL-90 profiles across six selected dimensions. The lower panel, B, plots the centred three-month moving mean of monthly PHQ-9 total scores from June 2023 through April 2026. The SCL-90 radar axes are somatization, obsessive-compulsive symptoms, depression, anxiety, phobic anxiety, and paranoid ideation, with a 1 to 5 scale. Seven profiles are dated May and December 2023, May and December 2024, May and December 2025, and May 2026. Later profiles are generally closer to the centre in anxiety-related dimensions. The PHQ-9 line varies over time: after falling during 2024 it rises again in 2025 and early 2026 before falling near the end. The vertical axis is PHQ-9 total score and the horizontal axis is assessment month. These records describe the follow-up period and do not estimate an effect of ChatGPT.}

\usepackage{xcolor}
\usepackage{array}
\usepackage{tikz}
\definecolor{FFCodeBg}{HTML}{E6FFCC}
\definecolor{ClinicalCodeBg}{HTML}{FFE8CC}
\definecolor{EpisodeCodeBg}{HTML}{DCEEFF}
\newcommand{\sourcecodebadge}[2]{\allowbreak\colorbox{#1}{\strut\textsf{\textbf{#2}}}}
\newcommand{\FFcode}[1]{\textbf{FF#1}}
\newcommand{\Dcode}[1]{\textbf{D#1}}
\newcommand{\Ccode}[1]{\textbf{C#1}}
\DeclareRobustCommand{\Ecode}[1]{\sourcecodebadge{EpisodeCodeBg}{E#1}}
\newcommand{\SSRI}{\textbf{SSRI}}
\newcommand{\SSRIs}{\textbf{SSRIs}}
\newcommand{\BZD}{\textbf{BZD}}
\newcommand{\BZDs}{\textbf{BZDs}}
\setcopyright{none}
\renewcommand\footnotetextcopyrightpermission[1]{}
\hypersetup{keeppdfinfo=true,pdflang={en-GB},pdfdisplaydoctitle=true,
  pdfauthor={Dongyijie Primo Pan; Pan Hui; Mirjana Prpa},
  pdfkeywords={panic disorder, ChatGPT, large language models, conversational agents, autoethnography, mental health, reassurance seeking, longitudinal qualitative research, trajectory-level safety}}

\begin{document}

\title[Can I Trust My Body?]{Can I Trust My Body? A Three-Year Autoethnography of ChatGPT’s Place in My Support System for Panic Attacks}

\author{Dongyijie Primo Pan}
\affiliation{%
  \institution{Computational Media and Arts, The Hong Kong University of Science and Technology (Guangzhou)}
  \city{Guangzhou}
  \country{China}}

\author{Pan Hui}
\affiliation{%
  \institution{Computational Media and Arts, The Hong Kong University of Science and Technology (Guangzhou)}
  \city{Guangzhou}
  \country{China}}
\affiliation{%
  \institution{The Hong Kong University of Science and Technology}
  \city{Hong Kong}
  \country{China}}

\author{Mirjana Prpa}
\affiliation{%
  \institution{Computational Media and Arts, The Hong Kong University of Science and Technology (Guangzhou)}
  \city{Guangzhou}
  \country{China}}

\renewcommand{\shortauthors}{Pan et al.}

\authorsaddresses{}

\begin{abstract}
People increasingly seek mental health support from large language models, yet little is known about their use across years of recurrent panic. We present a three-year analytic autoethnography of the first author's ChatGPT use while living with panic disorder, drawing on conversations, personal records, and accounts from friends or family members and professionals. Narrative analysis traces how my questions shaped ChatGPT's roles and how earlier experiences influenced later responses to symptoms. Familiar explanations could make sensations less frightening, while changed symptoms renewed fears of serious illness. During sudden panic, advice could be difficult to follow, and some replies prompted further checking. Conversations could end while symptoms, checking, or help-seeking continued. We propose trajectory-level safety during and after panic: usable advice (Fit), a stopping point for repeated checking and reassurance seeking (Closure), and useful understanding and human support that remain available over time (Continuity).
\end{abstract}




\maketitle
% Plain metadata avoids leaking CCS typesetting commands into the PDF Subject.
\hypersetup{pdfauthor={Dongyijie Primo Pan; Pan Hui; Mirjana Prpa},pdfsubject={Human-centered computing: Empirical studies in HCI; Natural language interfaces. Applied computing: Health informatics.}}

\section{Introduction}

Panic attacks are sudden periods of intense fear or discomfort, often accompanied by strong bodily sensations and fear of dying or losing control \cite{apa2022,nimh2025}. Panic disorder involves recurrent, unexpected attacks followed by persistent worry or changes in behaviour \cite{apa2022}. Unlike the ongoing worry associated with generalised anxiety disorder, panic attacks can bring a sudden sense of bodily danger \cite{apa2022,nimh2025}. Across 25 countries, 13.2\% of respondents reported a lifetime panic attack and 1.7\% met criteria for panic disorder \cite{dejonge2016}.

Chest pain, a racing heart, and breathlessness can feel like a heart attack and lead people to seek emergency care \cite{nimh2025,huffman2002}. When panic disorder goes unrecognised during these visits, people may leave without an explanation for their symptoms or a route into mental health care \cite{fleet1996,huffman2002}. Repeated visits also add to healthcare use and costs \cite{katon1996,chang2019}.

Established care includes cognitive behavioural therapy (CBT) and clinician-managed medication \cite{nice2020,papola2022,nimh2025,chawla2022}. In my experience, ChatGPT was immediately accessible when my usual clinician or someone I trusted was unavailable. I could describe what was happening, ask follow-up questions, and consider what to do next. Other studies have also documented people's use of general-purpose chatbots for emotional and decision support \cite{luo2025}, including the value they place on immediate access and control over disclosure \cite{purohit2026}.

Prior work has evaluated purpose-built chatbots for mental
health support, including Woebot \cite{fitzpatrick2017},
Tess \cite{fulmer2018}, and Wysa \cite{inkster2018}.
More recently, a four-week randomised trial evaluated
Therabot, a generative AI chatbot for depression, anxiety,
and eating-disorder risk \cite{heinz2025}.
For panic disorder, a preliminary four-week randomised trial
of CBT delivered by a scripted, non-LLM chatbot involving
41 patients reported a significant reduction in panic severity within the chatbot group, but not within a control group given a book on panic disorder \cite{oh2020}.

Panic attacks are hard to observe prospectively: repeated 24-hour ambulatory monitoring of 43 patients captured only 13 attacks across 1,960 hours \cite{meuret2011}. Questionnaires and interviews can document recalled symptoms and help-seeking around panic attacks \cite{katerndahl2002}. In this study, dated ChatGPT conversations also show the questions I asked and the replies I received as concerns developed.

Existing studies provide limited understanding of how people use a general-purpose chatbot on their own across years of recurrent panic. It remains unclear how their questions shape the help offered in its replies, what they do after reading those replies, and how earlier conversations and experiences shape their judgements when symptoms return and their choices about seeking help from ChatGPT, friends or family members, or clinicians.

We address this gap through a three-year analytic autoethnography of the first author's panic-related ChatGPT use, including reflection during stable periods \cite{anderson2006}. We use \textbf{episode} to mean an occasion when I experienced panic-like symptoms. A \textbf{trajectory} is the time-ordered account we reconstructed around an episode, linking bodily concerns, conversations, actions, support, and later reflection. Informed by Riessman's narrative approach \cite{riessman1993,riessman2008}, we compared 43 trajectories and examined selected conversation excerpts. We drew on dated conversations and personal records, with contextual accounts from friends or family members and professionals. We treat ChatGPT as one resource within a care network of people, practices, and services used to manage panic. We ask:

\begin{description}
  \item[\textbf{RQ1.}] What roles did ChatGPT take on in panic-related conversations, and how did these roles relate to my questions and descriptions?
  \item[\textbf{RQ2.}] How did conversations with ChatGPT unfold around bodily sensations and measurements, and what followed them?
  \item[\textbf{RQ3.}] How did I use ChatGPT alongside self-regulation, support from friends or family members, and professional care over the three years?
\end{description}

In the selected replies, we identified five recurring roles. My questions and descriptions shaped which role became prominent. Explanations could make familiar sensations less frightening, but advice could be difficult to use during panic, and some replies prompted further checking. Conversations could end while symptoms or help-seeking continued. Across three years, I described becoming more selective about advice, less fearful of prescribed medication, and less likely to call friends or family members during attacks, while continuing to draw on friends or family members and professional care. Unfamiliar symptoms could still renew uncertainty.

These findings inform three design aims: \textbf{Fit}, advice usable in the current state; \textbf{Closure}, a stopping point for repeated checking and reassurance seeking; and \textbf{Continuity}, useful understanding and human support available over time. Together, these guide our proposal for \emph{trajectory-level safety} during and after panic.

\textbf{Contributions.} We make three contributions to HCI:
\begin{enumerate}
\item An empirical account of ChatGPT's roles in panic-related conversations and the actions and concerns recorded around those conversations.
\item A three-year comparison of 43 reconstructed trajectories and contextual accounts, tracing changes in symptom interpretation, medication understanding, use of advice, and human support (Section~\ref{sec:findings-cross-rq-next-step}; Table~\ref{tab:longitudinal-comparison}).
\item A set of design proposals organised around Fit, Closure, and Continuity to guide evaluation of chatbot support during panic, beyond the conversation, and across recurrence (Section~\ref{sec:discussion-design}).
\end{enumerate}

\section{Related Work}

\subsection{Panic Attacks and Clinical Care}

An acute panic attack is an abrupt surge of intense fear or discomfort that peaks within minutes. Symptoms may include palpitations, breathlessness, chest discomfort, dizziness, numbness, derealisation, and fears of losing control or dying \cite{apa2022,nimh2025}. Some people with panic disorder also become caught in compulsive self-checking, such as repeatedly measuring their pulse or blood pressure to reassure themselves that nothing is wrong \cite{clark2009panic}. One common account describes a false alarm of defensive fight-or-flight systems \cite{gorman2000}; research also links panic to brain systems involved in fear and threat \cite{guan2024}.

Clinical care first excludes immediate physical danger \cite{huffman2002}. If no acute cause is found, guidance recommends explaining panic, reviewing previous treatment, and arranging follow-up \cite{nice2020}. Cognitive models describe panic as a cycle in which bodily sensations are catastrophically misinterpreted \cite{clark1986,beck1985}, a tendency supported by meta-analysis \cite{ohst2018}. These interpretations heighten arousal and prompt avoidance and safety behaviours such as reassurance seeking and repeated checking \cite{barlow1989}. Panic-focused CBT addresses these relations through cognitive restructuring and interoceptive or situational exposure \cite{barlow1989}, is recommended in clinical guidance \cite{nice2020,nimh2025}, and has meta-analytic support \cite{papola2022}.

Medication may complement psychotherapy. \SSRIs{} are maintenance treatments that may take weeks to work \cite{chawla2022}, whereas benzodiazepines (\BZDs{}) can quickly reduce panic symptoms and help an acute \textbf{episode} settle. They are usually prescribed for brief periods \cite{nimh2025}. \BZD{} risks include sedation, tolerance, and dependence; combining them with alcohol or other central nervous system depressants can lead to respiratory depression or death \cite{fda2020}. Clinical care therefore stretches over time, through assessment, explanation, practice, medication management, and follow-up, rather than resting on a single act of \textbf{reassurance}.

\textbf{Reassurance} is not clinically neutral. In cognitive accounts of health anxiety, it can briefly reduce fear while reinforcing the belief that the threat required checking, making further reassurance more likely \cite{salkovskis1986,warwick1990}. Although much of this evidence comes from health anxiety, the mechanism is relevant to panic because reassurance seeking and repeated checking can likewise function as safety behaviours within panic cycles \cite{clark1986,barlow1989}. Repeated checking follows a similar pattern, eroding rather than strengthening confidence that the threat has passed \cite{rachman2002}. Online searching can extend this cycle: searches may escalate common symptoms towards serious conditions \cite{white2009}, health anxiety is associated with more searching \cite{mcmullan2019} and searching with greater distress \cite{starcevic2013}, and symptom checkers often favour risk-averse triage advice \cite{semigran2015}. Conversational agents give this cycle a new setting. What an answer says matters, and so does how one response follows another, because that sequence can reinforce the cycle or interrupt it.

\subsection{Treatment Programmes and Self-Directed Chatbot Use}

Existing digital mental-health research often starts with a tool designed for a particular form of support. Woebot \cite{fitzpatrick2017}, Tess \cite{fulmer2018}, and Wysa \cite{inkster2018} offer selected therapeutic and self-management activities. Reviews examine symptom outcomes \cite{he2023}, empathic design \cite{sanjeewa2024}, and continued use \cite{jabir2024}. Panic-focused tools offer tasks such as a four-week CBT programme \cite{oh2020}, heart-rate biofeedback \cite{mcginnis2022}, diaries, psychoeducation, and exercises \cite{kimPanic2024}, or app-guided exposure \cite{guth2025}. Other work examines clinician-facing monitoring \cite{ko2025} and the quality of consumer apps \cite{vansinger2015}. These studies help establish how particular forms of support can be delivered and evaluated through digital tools.

Generative models can also be used for specified purposes, including public-health outreach \cite{jo2023}, cognitive restructuring \cite{sharma2024}, psychiatric journalling \cite{kim2024}, and a defined treatment programme \cite{heinz2025}. These studies explore conversational flexibility while identifying continuing needs for reliability, traceability, and oversight. A generative system can therefore still operate within a planned intervention.

Our study examines the first author's use of general-purpose ChatGPT outside a chatbot treatment programme. Questions about bodily sensations, reassurance, coping, medication, and seeking help arose during everyday life and recurrent \textbf{episodes}. The focus is how a person decides what help to seek from the chatbot as these concerns arise, and how that use fits alongside other support.

\subsection{Everyday Appropriation, Reliance, and Care}

Outside a treatment programme, an LLM chatbot has no fixed therapeutic role. Users decide what to disclose, trust, revisit, and stop. Qualitative studies report valued immediacy and non-judgement \cite{purohit2026}, privacy \cite{choi2025}, and perceived empathy \cite{wang2025,zheng2025}, alongside limits in crises and complex social situations \cite{purohit2026,choi2025}. The same system may become an explainer, reassurance provider, companion, or decision aid \cite{luo2025}.

For longer-term support, studies emphasise actionable information and privacy \cite{yoo2026}, personalisation and alliance \cite{xu2026}, and human connection \cite{yuan2026}. Fluent empathy may nevertheless miss relational \cite{vafafar2026} or cultural \cite{aldaweesh2026} circumstances. Constant availability, memory, and validation may also prolong avoidance or checking \cite{barkhuff2026}, deepen reliance on chatbot interpretations \cite{shen2026}, or reinforce harmful beliefs \cite{yang2026}. Frameworks for responsible design \cite{cooper2026} and for AI's role in mental-health apps \cite{anvari2026} raise the same concerns.

Two mechanisms in this literature matter especially for panic. An always-available and agreeable chatbot can become a new site for the reassurance seeking and repeated checking described above \cite{clark1986,barlow1989}, as reports of AI-directed reassurance seeking \cite{barkhuff2026} and epistemic rabbit holes \cite{shen2026} show. And a chatbot's role is not fixed within a conversation: the same exchange can move from explanation to reassurance to triage as the user adds information \cite{luo2025,purohit2026}. These mechanisms have received little longitudinal attention across recurrent \textbf{episodes}, particularly over years of use alongside friends or family members and professional care.

Two properties of current assistants help explain why that sequence matters. Sycophancy occurs when an assistant matches a user's stated view or revises a correct answer after pushback \cite{perez2023}. Preference training can encourage this behaviour \cite{sharma2024syco}; it became publicly visible when OpenAI withdrew a ChatGPT update for excessive agreeableness in April 2025 \cite{openai2025syco}. In mental-health scenarios the same tendency appears as validating a user's framing and responding inappropriately to crisis cues \cite{moore2025}. Safety guardrails, meanwhile, are most predictable at the extremes: across risk-graded suicide queries, major chatbots consistently declined the highest-risk questions and answered the lowest, but were inconsistent in between \cite{mcbain2025}. Panic conversations rarely carry those explicit crisis cues. They bring acute distress and medical uncertainty instead, so the question is how each successive response shapes checking and help-seeking.

\subsection{Care as Coordinated Work: Handoff and Infrastructure}
\label{sec:rw-care-work}

Care is rarely carried out by one person, in one place, or at one moment; it depends on coordination across people, settings, and time \cite{strauss1985}, much of it invisible articulation work \cite{star1999silence}. Clinical handoff makes this visible: rather than a single message, it is a recursive process of preparing, transferring, and re-checking information \cite{abraham2012}, with gaps emerging between phases \cite{tang2007}. Home care extends this coordination to relatives and paid carers who may understand the task differently \cite{bossen2013}, while chronic care relies on arrangements patients assemble from medication, devices, routines, and calls \cite{langstrup2013}.

These arrangements form the infrastructure through which care continues across \textbf{episodes}. Star and Ruhleder describe infrastructure as relational: embedded in practice, learned through participation, and often visible only when it breaks down \cite{star1996,star1999}; Bowker and Star's \emph{infrastructural inversion} similarly brings such normally invisible supports into view \cite{bowker1999}. We use these ideas in two ways. First, handoff to friends or family members or professionals requires coordination that a chatbot may help prepare but cannot itself complete. Second, the explanations, routines, and contacts available after one \textbf{episode} form a personally assembled care infrastructure for the next.

\subsection{Longitudinal Health Autoethnography in HCI}

Health technologies may change purpose through continued use: a device may record symptoms, become a source of reassurance or a medium for clinical communication, and later be set aside when it adds uncertainty. Short deployments and retrospective interviews often miss these shifts. Autoethnography connects bodily experience, later reflection, and care \cite{ellis2011,anderson2006,chang2016}; in HCI, autobiographical design \cite{neustaedter2012,desjardins2018} and a growing body of autoethnographies \cite{kaltenhauser2024} study technology use from the inside. Studies of Long COVID \cite{homewood2023}, menstrual tracking \cite{homewood2020}, pregnancy \cite{gamboa2023}, glucose monitoring \cite{cgm2025}, migraine \cite{sefidgar2024}, diabetes \cite{james2025}, self-tracking metrics \cite{loerakker2025}, and polycystic ovary syndrome \cite{kang2025} show how tracking practices, interpretations, and care relations evolve. Homewood shows how bodily change can alter a tracker's purpose \cite{homewood2023}. We extend that longitudinal lens to generative dialogue: how conversation reorganises bodily interpretation, checking, and movement into care.

\section{Method}

\subsection{Study Design}

We conducted a longitudinal analytic autoethnography of the first author's use of ChatGPT from May 2023 to July 2026, with the first author as participant--researcher. Over these three years the service changed its underlying models, memory features, and safety behaviour. ChatGPT had entered everyday life through work, practical problem solving, and health information seeking before it became involved in panic-related situations. We followed when he turned to ChatGPT for help with panic, what help he sought, and how he used it alongside support from friends or family members and professionals. Autoethnography suited interactions that were private, episodic, and difficult to observe prospectively \cite{ellis2011,anderson2006,chang2016}.

An \textbf{episode} was an occasion when the first author experienced panic-like symptoms. A \textbf{trajectory} was the time-ordered account we constructed around that \textbf{episode} from the available records. The \textbf{trajectory} was our primary analytic unit. Each \textbf{trajectory} began with an initiating bodily concern and connected what the participant knew, what was recorded at the time, what was done, how the state changed, who helped, and what was reflected on later. Where ChatGPT was involved, the \textbf{trajectory} included what it was told and what it replied. \textbf{Trajectories} could span anticipatory, acute, recovery, and reflective states without fixed durations.

A \emph{stable period} was a time when no panic-like symptom \textbf{episode} was underway. Health worries or other discomfort could still be present. Conversations during these periods included medication questions, planning care, and reflecting on earlier experiences.

\begin{table*}[!htbp]
  \centering
\caption{Episode and source labels used throughout the manuscript. Each E label identifies one panic-like symptom episode; C, D, and FF labels identify the counsellor, psychiatrist, and friends or family members.}
  \label{tab:source-labels}
  \small
  \begin{tabular}{@{}lp{0.72\linewidth}@{}}
    \toprule
    Label & What it identifies \\
\midrule
\Ecode{01}--\Ecode{43} & Panic-like symptom \textbf{episodes}, numbered from 01 to 43. Each number identifies one \textbf{episode}; its \textbf{trajectory} connects the related records. \\
\cmidrule(lr){1-2}
    ChatGPT & OpenAI's general-purpose LLM chatbot \\
    \cmidrule(lr){1-2}
    \Ccode{1} & Psychological counsellor \\
    \cmidrule(lr){1-2}
    \Dcode{1} & Psychiatrist involved in the gut--brain axis research programme \\
    \cmidrule(lr){1-2}
    \FFcode{1} & Friend or family member (specific relationship withheld) \\
    \cmidrule(lr){1-2}
    \FFcode{2} & Friend or family member (specific relationship withheld) \\
    \bottomrule
  \end{tabular}
\end{table*}

We use the umbrella category \emph{friends or family members} to protect specific relationships. \FFcode{1} and \FFcode{2} identify two distinct individuals within this category; the labels do not specify whether either person is a friend or a family member.

\subsection{Researcher--Participant Positionality}

The first author is an HCI researcher in his twenties. His digital and health-data literacy, self-tracking experience, and access to monitoring devices and professional care shaped both system use and analysis. Research training also shaped how he prompted and what he retained. He often assembled symptoms, measurements, and earlier explanations into explicit problems.

\textbf{Vignette}: \textit{Before the study, while he was in high school, the first author experienced numbness in the hands and feet and vision going dark. An emergency assessment found no abnormality, but the event was not framed as a possible panic attack. Similar symptoms recurred, leading him to learn about panic independently and to enter psychiatric care, where he reports receiving a diagnosis of panic disorder. Subsequent years included recurrent attacks, periods of improvement, and medication-free periods. During the study, clinician-directed care included maintenance treatment with a selective serotonin reuptake inhibitor (\SSRI{}), as-needed benzodiazepine (\BZD{}) use when acute \textbf{episodes} were otherwise difficult to manage, and routine physical examinations. Much of his panic knowledge and care practice therefore predated ChatGPT.}

Being both researcher and patient gave close access to bodily uncertainty and help-seeking. During \textbf{episodes}, he asked to be told that he was safe. Between \textbf{episodes}, he assembled symptoms, readings, and history as problems. To examine how his own participation shaped the exchanges, he used source-labelled chronology, reflexive memos, and counterexamples. The selected conversation excerpts preserve turns in which his own contribution visibly changed the reply.

\subsection{Corpus and \textbf{Trajectory} Construction}

The archive comprised retained ChatGPT and social-media chats; reflexive records and records involving friends or family members; personal health and clinical records; and a timeline assembled from dated materials, including photographs and timestamps. The ChatGPT sub-corpus was identified by searching the first author's ChatGPT history for panic-related sensations and concerns, including palpitations, fear of dying, and benzodiazepines, together with participant-specific terms withheld as private health information. A conversation was one archived ChatGPT thread, including all retained messages in their original order. A turn was one message written by either the participant or ChatGPT. A reply was a turn written by ChatGPT.

The sub-corpus contained 87 conversations about panic and related health concerns (3,328 turns; approximately 610,000 Chinese characters). These conversations included support during panic-like symptoms, related health questions, and reflection during stable periods. Conversations that broke off without resolution were retained and coded as they stood.
The archive documented 43 panic-like symptom \textbf{episodes} between July 2023 and June 2026. We constructed one \textbf{trajectory} for each, giving 43 \textbf{trajectories}. The identifiers \Ecode{01}--\Ecode{43} refer to these \textbf{episodes} throughout the paper and figures. We linked records by event and date. Each \textbf{trajectory} required an initiating panic-related concern and at least one subsequent action, change in state, or later reflection; all cases meeting these criteria were retained. For example, \Ecode{30} names the \textbf{episode} of dizziness and fear in June 2025. Its \textbf{trajectory} connects that experience with the emergency call, a roommate's accompaniment, clinical assessment, and later monitoring. ChatGPT use was recorded in 31 \textbf{episodes} and not in 12. One \textbf{trajectory} could connect several conversations with other records. Conversations outside these \textbf{episodes} provided context. Figure~\ref{fig:panic-episode-diary-map} locates the 43 \textbf{episodes}.

A \textbf{conversation excerpt} was a set of consecutive turns selected from one conversation and kept in their original order for close reading. A \emph{contextual account} was a source-labelled diary entry, memo, or attributed account from a friend or family member or a professional describing the surrounding \textbf{episode}; such accounts provided context alongside the conversation excerpt. A \emph{stable-period conversation} took place during such a period and was analysed as contextual material outside the reconstructed \textbf{trajectories}. Table~\ref{tab:material-boundaries} summarises what each material was used to show.

\begin{table*}[!htbp]
  \centering
  \caption{Analytic use and handling of each source material.}
  \label{tab:material-boundaries}
  \small
  \setlength{\tabcolsep}{4pt}
  \renewcommand{\arraystretch}{1.12}
  \begin{tabular}{@{}>{\raggedright\arraybackslash}p{0.20\linewidth}
    >{\raggedright\arraybackslash}p{0.30\linewidth}
    >{\raggedright\arraybackslash}p{0.42\linewidth}@{}}
    \toprule
    Material & Used here to show & Source handling \\
    \midrule

    87 retained ChatGPT conversations &
    Turn order, questions, replies, and thread endings; a selected set of 792 replies was coded for roles and role changes. &
    Each thread retained as a conversation; three conversation excerpts read closely. \\

    \cmidrule(lr){1-3}

    43 \textbf{episode} sheets (one per \textbf{trajectory}) and the longitudinal timeline &
    Source-linked \textbf{episode} context, actions, changes in state, and movement into other care. &
    Records linked by event and date across conversations and other sources; source gaps and retrospective elements labelled. \\

    \cmidrule(lr){1-3}

    Stable-period conversations &
    Anticipatory planning, medication questions, and reflection when no \textbf{episode} was underway. &
    Read as context for use between \textbf{episodes}, separately from the reconstructed \textbf{trajectories}. \\

    \cmidrule(lr){1-3}

    Other dated records &
    Surrounding measurements, actions, care encounters, and later interpretation. &
    Linked to \textbf{episode} sheets with dates and provenance; kept distinct from archived dialogue. \\

    \cmidrule(lr){1-3}

    Accounts from friends or family members and professionals &
    Attributed descriptions of change, support, and responsibility. &
    Retained as attributed contextual accounts. \\

    \cmidrule(lr){1-3}

    PHQ-9 and SCL-90 records &
    Descriptive context at monthly and approximately semiannual timescales. &
    Displayed descriptively at the programme's assessment intervals (Appendix~\ref{app:longitudinal-scales}). \\

    \bottomrule
  \end{tabular}
\end{table*}

\subsection{Analysis}

The materials were analysed through case-centred thematic narrative analysis, informed by Riessman \cite{riessman1993,riessman2008}. We performed three operations: linked records to describe what happened around each \textbf{episode}; compared \textbf{trajectories} to identify what recurred, changed, or carried into later \textbf{episodes}; and coded ChatGPT replies and read conversation excerpts to examine roles, role changes, and endings. The first author constructed all \textbf{trajectories} and did all coding, translation, and analysis. The two coauthors advised on study design, theoretical framing, evidentiary boundaries, and manuscript revision, but did not inspect or code the private corpus. \textbf{Episode} sheets, codes, and memos were written in Chinese, and all quoted excerpts were translated into English by the first author from written Chinese or transcribed Mandarin; the Chinese materials remain the controlling sources.

\textbf{Episode} boundaries were traced across all sources, not from chats alone. Each \textbf{trajectory} was then written up as an \textbf{episode} sheet recording provenance, context, sensations or measurements, available knowledge, ChatGPT involvement, subsequent action, human support, and later reflection, with retrospective elements and source gaps marked. Short process codes, such as \emph{reporting a new metric} and \emph{contacting friends or family members}, indexed actions and transitions, and comparative memos traced turning points, counterexamples, role changes, and the work a reply made relevant next. In the analysis, we distinguished what the first author already knew before a conversation, what ChatGPT explained or suggested during it, and what the first author came to understand when reflecting on the experience afterwards. For the longitudinal comparison, we juxtaposed earlier and later accounts of symptom interpretation, medication understanding, help-seeking, and use of advice, distinguishing dated exchanges from retrospective descriptions of change.

To answer RQ1, we developed five roles through comparative memoing: \textbf{Explainer}, \textbf{Reassurance provider}, \textbf{Action guide}, \textbf{Safety checker}, and \textbf{Interrupter of compulsive self-checking}. These describe recurring forms of panic-related support. We assigned a \emph{primary role} when one predominated across a complete reply. The first author screened all 87 conversations in full and retained 792 replies about panic or related bodily symptoms, excluding unrelated exchanges within the same threads. All 792 replies remained in the analysis: 71 combined roles with no clear primary; 59 contained panic-related material outside the five roles, including privacy-sensitive personal interpretations; and 26 could not be classified in context, including replies that hallucinated context or invoked irrelevant memories. Poor quality alone did not prevent role assignment. We counted a switch when successive replies with clear primary roles differed within the same conversation, skipping replies without one. Changes in topic or content alone did not count. Role and switch counts were descriptive. Close readings and comparisons across \textbf{trajectories} examined conversation endings and available records of subsequent checking and other actions. We also examined \textbf{stopping rules}, which stated when to stop an action and how to decide whether to resume it, and observation windows, which specified when to review the situation.

Three conversation excerpts were selected for contrast: \Ecode{15}, where an unusable next step was followed by safety checking; \Ecode{24}, where repeated measurement was followed by a proposed stopping rule; and \Ecode{06}, a boundary case with overlapping roles and no defensible switch. Selection required retained participant turns and the following ChatGPT replies, so that any claimed role change could be read against what had been typed, together with contextual records for the \textbf{episode} (Table~\ref{tab:three-case-comparison}). Figure~\ref{fig:selected-reply-sequences} gives condensed English translations: wording within each displayed turn was shortened, turn order was kept, and nothing was added. Comparisons across conversation excerpts and \textbf{trajectories} informed the concepts developed in the Discussion.

\subsection{Contextual Materials and Professional Consultations}

The first author also took part in a national gut--brain-axis research programme: a three-month double-blind intervention from May to August 2023, followed by three years of follow-up with monthly PHQ-9 questionnaires \cite{kroenke2001}, telephone follow-ups, semiannual SCL-90 assessments \cite{derogatis1973}, and annual electroencephalography and functional magnetic resonance imaging. The research team authorised use of the participant-accessible PHQ-9 and SCL-90 records and the consultation with \Dcode{1} only. \Dcode{1} recommended six SCL-90 dimensions for descriptive display as those most relevant to the participant's panic-related presentation: somatisation, obsessive-compulsive symptoms, depression, anxiety, phobic anxiety, and paranoid ideation (Appendix~\ref{app:longitudinal-scales}). The selection was not based on their observed patterns. Separately, after the treating physician assessed the condition as stable, routine \SSRI{} maintenance was paused under clinical guidance from late 2024 until the June 2025 emergency visit (\Ecode{30}), and resumed the next day at psychiatric follow-up. The plan permitting as-needed \BZD{} use remained in place throughout.

For professional context, the first author consulted \Ccode{1}, a psychological counsellor seen monthly throughout the study, and \Dcode{1}. Counselling-service policy prohibited recording the consultation with \Ccode{1}; with consent it was documented in contemporaneous notes, so \Ccode{1} is paraphrased rather than quoted. The consultation with \Dcode{1} was recorded and transcribed. Contextual material from friends or family members comprised seven audio-recorded calls with \FFcode{1} and \FFcode{2} about perceived change across three years, recurrent attacks, reassurance seeking, emergency decisions, and coping.

\subsection{Ethics and Institutional Review}

Autoethnographers must protect both implicated others and themselves \cite{cooperLilyea2022}. Because the calls with \FFcode{1} and \FFcode{2} and the consultations with \Ccode{1} and \Dcode{1} included jointly experienced \textbf{episodes} and private actions, we obtained institutional review board (IRB) approval for their use with appropriate consent and safeguards.

\section{Findings}

The Findings, and the first-person passages of the Discussion, use the first
person singular for lived experience and first-author interpretation; ``we''
refers to the authors' shared analytic and design claims.

\noindent\textbf{Longitudinal orientation: improvement without disappearance.}
Periodic PHQ-9 and SCL-90 records from the follow-up programme provide descriptive context at monthly and approximately semiannual timescales (Appendix~\ref{app:longitudinal-scales}). Across the seven SCL-90 assessments, the anxiety-related dimensions were broadly lower in the later records. In the interview, the psychiatrist (\Dcode{1}) described the later records as broadly stable and said that ``overall, the recovery was still quite good.'' The smoothed monthly PHQ-9 record remained uneven, with renewed increases in late 2025 and early 2026. In our conversation, \Dcode{1} explained that PHQ-9 could reflect my overall emotional state during that period. In these terms, the lower later anxiety-related scores coexisted with fluctuations in my broader emotional state. Although the scale records suggested some improvement, \Dcode{1} emphasised that the mental status examination, based on conversation and observation, remained psychiatry's ``most fundamental method.''

\begin{figure*}[!tp]
  \centering
  \IncludePDFWithOwnGroup[width=\textwidth,height=.76\textheight,keepaspectratio]{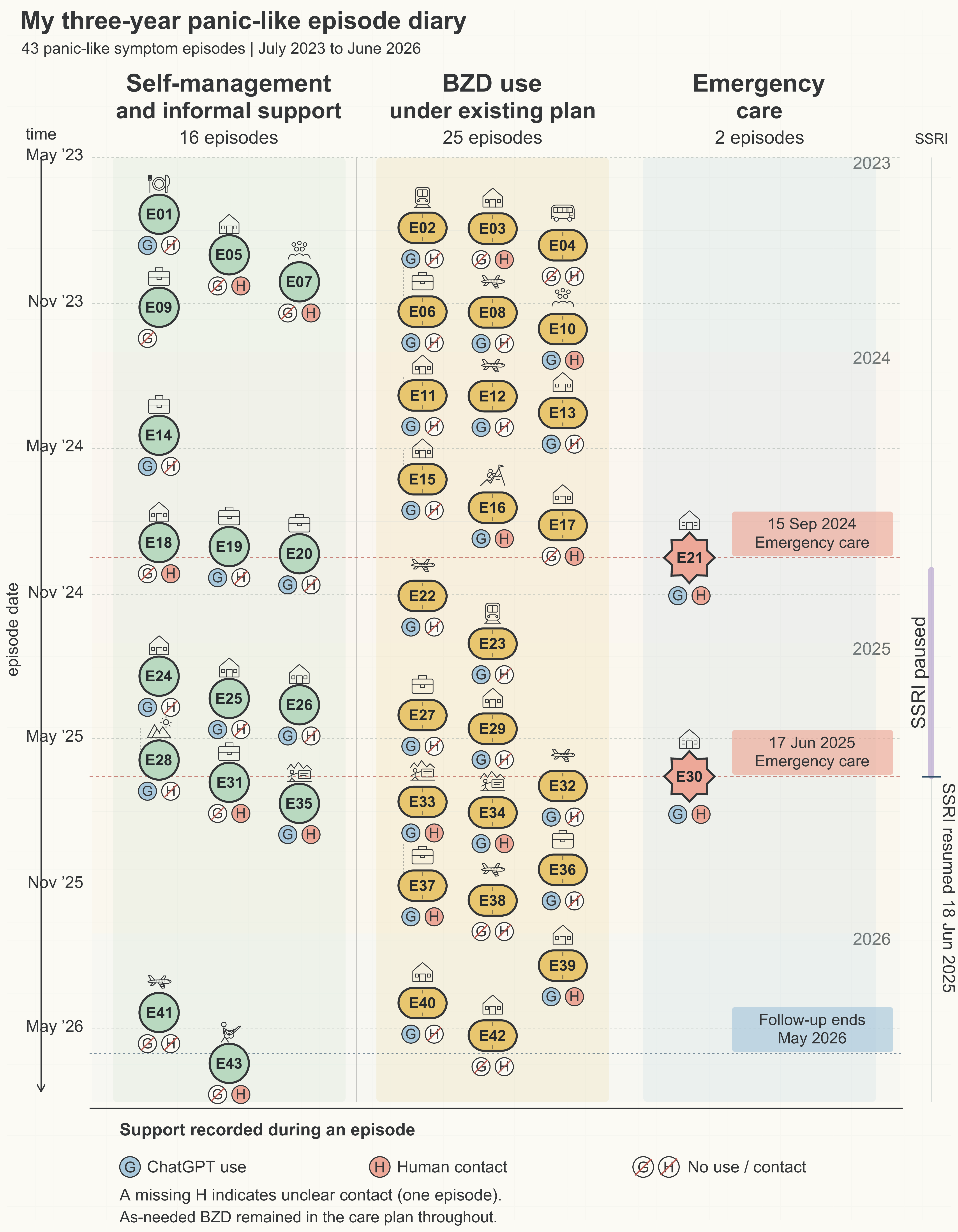}
  \caption{Three-year diary map of 43 panic-like symptom episodes (July 2023--June 2026). Each main symbol contains one episode ID, E01--E43; each ID refers only to that symbol. Read the timeline from top to bottom. Columns group episodes by the highest recorded care step: self-management and informal support without BZD or emergency care (16; green circles), BZD use under the existing care plan without emergency care (25; yellow pill shapes), and emergency care (2; star shapes). Small drawings above the symbols indicate episode settings, such as home, work, rail travel, and air travel. G and H badges below the symbols indicate ChatGPT use and contact with another person. A slash indicates non-use or non-contact; a missing H indicates unclear contact. Dashed connectors link displaced symbols to their calendar dates; spacing does not indicate trajectory duration. The right-hand rail shows changes in routine SSRI maintenance. The May 2026 line marks the end of quantitative follow-up; the episode diary continues into June. BZD = benzodiazepine; SSRI = selective serotonin reuptake inhibitor.}
  \Description{\DiaryAlt}
  \label{fig:panic-episode-diary-map}
\end{figure*}

Improvement also coexisted with recurrence. We reconstructed 43 \textbf{trajectories}, one for each documented panic-like symptom \textbf{episode} (Figure~\ref{fig:panic-episode-diary-map}).
The \textbf{episodes} I would describe to a clinician as full panic attacks
became much less frequent over the three years, whereas panic-like symptoms
continued to occur. The highest care step recorded was \BZD{} use under the existing clinician-directed plan, without emergency care, in 25 \textbf{episodes}, and emergency care in two. ChatGPT appeared in 31, and contact with another person
was recorded in 16. The same \textbf{episode} could include one or more forms of support.

Contextual accounts described the same change with different criteria. My counsellor (\Ccode{1}) described me as having more energy and being
better able to handle difficulties and respond to panic than at our
first meeting. Two friends or family members (\FFcode{1} and \FFcode{2}) located change in how
support was needed: familiar \textbf{episodes} became patterns I could navigate without
always requiring immediate repeated confirmation, although they still helped when symptoms felt harder to cope with or different from what I had experienced before. My own assessment was
similar. Panic became easier to recognise and carry, yet a recurrence could
still send me back to repeated reassurance, checking, rescue medication, or
emergency care. Improvement did not mean that \textbf{episodes} disappeared. It meant an uneven
change in what I could understand and do, in which I drew on different sources of support during an \textbf{episode} with ChatGPT becoming an important part of my care network. %

\begin{samepage}
\subsection{ChatGPT Shifted Roles as the Help I Needed Changed (RQ1)}
%RQ1: What roles did ChatGPT take on in panic-related conversations, and how did these roles relate to my questions and descriptions?
\label{sec:findings-state-dependent-support}

Across the 792 analysed replies about panic and related bodily symptoms, I identified five roles: The examples below paraphrase exchanges in Figure~\ref{fig:selected-reply-sequences}:
\begin{itemize}
\item \textbf{Explainer} offered a possible reason for a symptom: in \Ecode{06}, ChatGPT linked numb hands and feet to rapid breathing and explained how this could lower carbon dioxide and make the blood more alkaline (respiratory alkalosis), causing tingling or numbness \cite{lewis2026alkalosis}.
\item \textbf{Reassurance provider} tried to ease immediate fear: in \Ecode{24}, it said that a single blood-pressure reading alone did not establish an emergency.
\item \textbf{Action guide} suggested something to do: in \Ecode{15}, it asked me to sit or half-recline and slow my exhale.
\item \textbf{Safety checker} checked whether outside help was needed: in \Ecode{15}, it asked about warning signs and later advised urgent care.
\item \textbf{Interrupter of compulsive self-checking} addressed checking itself: in \Ecode{24}, it asked me to stop measuring blood pressure for reassurance.
\end{itemize}
\end{samepage}
A reply could combine roles, as happened in \Ecode{06}. A primary role was the role that predominated in one complete reply; it did not mean the role at the start of the conversation. A switch occurred when successive replies with a clear primary role differed.

\noindent\textbf{Changes in ChatGPT's primary role.}
Figure~\ref{fig:role-transitions} shows 222 changes in ChatGPT's primary role within the 792 analysed replies.
Rows indicate the role before a change and columns the role after it;
each cell gives the number of such changes.
Within each conversation, we compared ChatGPT messages assigned a clear primary role in their original order, skipping messages without one. The compared messages were therefore not necessarily adjacent in the original conversation.

Among replies assigned a clear primary role, \textbf{Explainer} and \textbf{Safety checker} were the most frequent. When the primary role changed from \textbf{Reassurance provider}, the most frequent next roles were \textbf{Action guide} and \textbf{Explainer}.
Changes from \textbf{Safety checker} to \textbf{Reassurance provider} were recorded 27 times, compared with three in the reverse direction. This asymmetry describes the screened material under the skipping rule; it does not establish a general tendency of the model.

In replies coded as \textbf{Interrupter of compulsive self-checking}, ChatGPT identified repeated measurement or requests for confirmation and proposed a stopping point. This label describes ChatGPT's proposal, not evidence that I stopped checking. When the primary role changed from \textbf{Interrupter of compulsive self-checking}, it most often changed to \textbf{Action guide} (11 recorded switches). No switches from \textbf{Interrupter of compulsive self-checking} to \textbf{Safety checker} were recorded in the coded sequences.

The close readings below examine two less frequent role changes: \textbf{Action guide} to \textbf{Safety checker} in \Ecode{15}, and \textbf{Reassurance provider} to \textbf{Interrupter of compulsive self-checking} in \Ecode{24}. These changes were recorded two and seven times, respectively, across the coded conversations. Each excerpt preserves my message preceding the change, showing what I reported before ChatGPT's response shifted to a different primary role.

\begin{figure*}[t]
\centering
\par% Included by the Findings figure; values preserved from the original matrix.
\begingroup
\sffamily\small
\setlength{\tabcolsep}{4pt}
\renewcommand{\arraystretch}{1.45}
\definecolor{matrixink}{HTML}{5B7C99}
\definecolor{matrixcoral}{HTML}{E85D47}
\newcommand{\switchcell}[2]{\colorbox{matrixink!#1}{\makebox[1.7em]{\strut #2}}}
\newcommand{\markedcell}[2]{\fcolorbox{matrixcoral}{matrixink!#1}{\makebox[1.7em]{\strut #2}}}
\begin{minipage}[t]{0.56\linewidth}
\centering
\begin{tabular}{@{}lccccc@{}}
 & \multicolumn{5}{c}{\textbf{To role (after the switch)}}\\
\textbf{From role (before)} & \textbf{E} & \textbf{R} & \textbf{A} & \textbf{S} & \textbf{I}\\
\hline
\textbf{E} & --- & \switchcell{60}{31} & \switchcell{37}{19} & \switchcell{41}{21} & \switchcell{20}{10}\\
\textbf{R} & \switchcell{35}{18} & --- & \switchcell{39}{20} & \markedcell{7}{3} & \switchcell{15}{7}\\
\textbf{A} & \switchcell{24}{12} & \switchcell{0}{0} & --- & \switchcell{5}{2} & \switchcell{17}{8}\\
\textbf{S} & \switchcell{30}{15} & \markedcell{53}{27} & \switchcell{13}{6} & --- & \switchcell{5}{2}\\
\textbf{I} & \switchcell{11}{5} & \switchcell{11}{5} & \switchcell{22}{11} & \switchcell{0}{0} & ---\\
\end{tabular}
\end{minipage}\hfill
\begin{minipage}[t]{0.42\linewidth}
\begin{tabular}{@{}l>{\raggedright\arraybackslash}p{0.65\linewidth}r@{}}
 & \textbf{Primary role} & \textbf{Replies}\\
E & \textbf{Explainer} & 207\\
R & \textbf{Reassurance provider} & 114\\
A & \textbf{Action guide} & 63\\
S & \textbf{Safety checker} & 155\\
I & \textbf{Interrupter of compulsive self-checking} & 97\\
\end{tabular}
\par\smallskip
\parbox{\linewidth}{\footnotesize Cells count switches; darker shading means more switches.
Coral outlines compare Safety checker $\rightarrow$ Reassurance provider (27) with the reverse (3).
Reply totals describe role assignments, not switches.}
\end{minipage}
\endgroup
\par
  \caption{Changes in primary role between successive replies with a clear
  primary role within the 792 analysed replies (row: role before;
  column: role after; 222 switches). Of the 792 replies, 636 received a clear
  primary role, 71 combined roles without a clear primary, 59 contained other
  panic-related content outside the five-role analysis, and 26 had no
  defensible role assignment in context (see Methods). Replies without a clear
  primary role were omitted from transition counting rather than treated as
  sequence breaks. The separate role key gives the total number of replies assigned each primary role. The coral outline marks the asymmetry between safety checking
  followed by Reassurance and Reassurance followed by Safety checking.}
  \Description{\RoleMatrixAlt}
  \label{fig:role-transitions}
\end{figure*}

\noindent\textbf{\Ecode{15}: from Action guide to Safety checker.}
\Ecode{15} (Figure~\ref{fig:selected-reply-sequences}, left) began with me telling ChatGPT that I felt I was going to die and that even slight movement made my heart race. The first reply suggested sitting or half-reclining and trying a breathing and grounding step. Although it also mentioned warning signs, I coded its primary role as \textbf{Action guide} because it mainly offered a coping step. I replied that I could not control my breathing and that my heart rate was still rising. The next reply focused on whether my symptoms required medical assessment: it stated that chat alone could not rule out other causes and described when urgent care would be needed. I coded this reply as \textbf{Safety checker}. After I reported raised blood pressure and feeling faint when standing, the final reply advised me to contact someone and seek emergency care.

\noindent\textbf{\Ecode{24}: from Reassurance provider to Interrupter of compulsive self-checking.}
\Ecode{24} (Figure~\ref{fig:selected-reply-sequences}, centre) began with a
blood-pressure reading I wanted ChatGPT to tell me if it was safe or dangerous.
The first reply was primarily from the role of  \textbf{Reassurance provider}. I measured again and reported the
number; the next reply still answered the present risk, but began to ask why I
was measuring again. I then reported that the reading was ``finally'' 125/79. The next reply treated this as the point at which I allowed myself to feel safe and explicitly asked me to stop checking. Here the primary role was clearly \textbf{Interrupter of compulsive self-checking}. I subsequently disclosed more than 30 measurements in half an hour, and the final reply asked me to stop and follow my doctor's advice on future measurement. The change developed as my reports made the checking pattern explicit.

\noindent\textbf{\Ecode{06}: overlapping roles, uncertain boundary.}
In \Ecode{06} (Figure~\ref{fig:selected-reply-sequences}, right), a racing heart had
raised the fear of a heart attack; when I added numbness, I began to fear a stroke. The ChatGPT replies kept offering a mechanism, reassurance, action steps, and
warning signs together, and explanation, reassurance, and safety checking
stayed intertwined. The feared condition and the reply content changed, but no
single role became clearly primary, so I labeled the excerpt as one without a
switch.
Table~\ref{tab:three-case-comparison} compares the three conversation excerpts with their
contextual records.

\begin{figure*}[tp]
  \centering
  \par\input{panic_autoethnography_chi_sections/figures/analytic/conversation-excerpts-reviewed.tex}\par
  \caption{Condensed English translations of selected excerpts from three
  conversations (\Ecode{15}, \Ecode{24}, and \Ecode{06}). Bold role labels and arrows mark where a change becomes clear; \Ecode{06} retains overlapping roles without an asserted switch. The displayed turn order is preserved and wording is condensed. The archived Chinese-language exchanges
  remain the controlling records. Purple boxes are participant turns; blue-grey
  marks a ChatGPT reading of the current state; teal marks reframing or a next
  step; coral marks a safety or interaction boundary. These colours are
  first-author analytic annotations.
  Vertical order shows observed turn order. The rebreathing advice recorded in
  the \Ecode{06} panel conflicts with current guidance \cite{anzcor2026}.}
  \Description{\ConversationAlt}
  \label{fig:selected-reply-sequences}
\end{figure*}

\begin{table*}[!htbp]
  \centering
  \caption{Three-case comparison of conversation excerpts and contextual records.
  The final column states what the retained record shows for each case.}
  \label{tab:three-case-comparison}
  \small
  \setlength{\tabcolsep}{3pt}
  \renewcommand{\arraystretch}{1.14}
  \begin{tabular}{@{}>{\raggedright\arraybackslash}p{0.055\linewidth}
    >{\raggedright\arraybackslash}p{0.15\linewidth}
    >{\raggedright\arraybackslash}p{0.19\linewidth}
    >{\raggedright\arraybackslash}p{0.17\linewidth}
    >{\raggedright\arraybackslash}p{0.15\linewidth}
    >{\raggedright\arraybackslash}p{0.215\linewidth}@{}}
    \toprule
    Case & Initial uncertainty & Changes in ChatGPT's role & What remained to be done & Who was involved & What the record shows \\
    \midrule
    \Ecode{15} & A racing heart with slight movement felt like imminent bodily danger.
    & Action guidance gave way to safety checking after I reported an unusable
    breathing step and added bodily information.
    & Respond to changing sensations and decide how to obtain help outside the chat.
    & ChatGPT asked and proposed; I reported and chose; in-person help was named.
    & Selected turns show an urgent-care proposal; the inventory records
    \BZD{} use and no emergency attendance. \\
    \cmidrule(lr){1-6}
    \Ecode{24} & Raised readings and dizziness recalled an earlier frightening
    blood-pressure reading.
    & Reassurance gave way to interrupting repeated checking as my reports made
    the pattern explicit.
    & Put the proposed pause into practice and decide whether to use the future
    checking rule.
    & ChatGPT named the pattern and proposed a rule; I would have to enact it.
    & Reported measurements and a stopping proposal; the conversation excerpt ends with the proposal. \\
    \cmidrule(lr){1-6}
    \Ecode{06} & A racing heart raised fear of a heart attack; numbness then raised
    fear of stroke.
    & Explanation, reassurance, action guidance, and safety questions overlapped;
    no clear primary-role switch was identified within the excerpt.
    & Manage immediate distress and interpret similar sensations when they
    recur; action safety remains a separate concern.
    & ChatGPT offered accounts and actions; I supplied symptoms and interpreted them.
    & Selected turns support the close reading; later reflection reports less
    troubling numbness. \\
    \bottomrule
  \end{tabular}
\end{table*}

\noindent\textbf{In anticipatory states, ChatGPT could also plan how care would be reached.}
In a stable-period conversation (July 2026) about returning to a psychiatric
hospital I had visited many times, ChatGPT first treated the route as a problem of time, cost,
and transfers. After I explained that a 40-minute metro journey could become
difficult during anxiety, the plan became state-dependent: ``On a stable day:
high-speed rail plus metro. On a day with clear pre-attack signs: take a
taxi.'' The exchange shows an anticipatory role absent from the acute conversation excerpts:
while I could still plan, ChatGPT helped set the conditions under which
professional care would feel reachable.

\noindent\textbf{Support varied between \textbf{episodes} across all three years.}
These exchanges show how support changed within conversations. The inventory
of 43 \textbf{trajectories} shows how it varied between \textbf{episodes}: not one kind of
support steadily replaced another over the three years. \Ecode{01} and
\Ecode{02} both recorded ChatGPT use without contact with another person, yet only
\Ecode{02} included \BZD{} use; much later, \Ecode{33}--\Ecode{35} fell within one short period and
all recorded ChatGPT and support from friends or family members, with \BZD{} use in the first two
and not the third. Medication, contact, and escalation describe what entered an \textbf{episode}.

The \textbf{episode} memos connect these patterns to what I could do in the moment. In \Ecode{15}, my symptoms felt very severe. I could still open ChatGPT and read its breathing instruction, but I typed back that I could not do it. Being able to continue the conversation did not mean that I could begin the suggested step while I felt in danger. Other memos showed a similar difference. Putting a poor sleep score in context helped when I otherwise felt stable; similar reassurance did little when I already felt bodily unease and expected a bad day. Afterwards, I could understand an explanation that had been difficult to draw on during panic. These accounts distinguish what a reply offered, what I could use at that moment, and what I did next.

\subsection{Making Sense of Symptoms Could Ease Fear or Restart Checking (RQ2)}
%RQ2: How did conversations with ChatGPT unfold around bodily sensations and measurements, and what followed them?
\label{sec:findings-bodily-legibility}

Here, outcomes refer to what the records show about how I understood symptoms, what I checked or did, and whom I contacted. %
A racing heart could recall a memory of an
earlier emergency visit. A raised blood pressure reading could evoke another \textbf{episode} from my  memory. Symmetric numbness could suggest one acute danger and then another.
I interpreted these sensations and readings through familiar patterns and earlier experiences before asking ChatGPT about them.
The conversation then shaped how I described the sensations, which readings I returned to, and what I checked next. I was asking both \emph{What does this mean now?} and
\emph{Is the earlier danger happening again?}

\noindent\textbf{More information could ease my fear or give me more reasons to worry.}
In one retrospective memo, I described an unusual watch reading alongside the readings immediately before and after it. Those surrounding readings were within my usual range. ChatGPT suggested looking at these readings together. Doing so made the unusual number less alarming to me. \Ecode{21} was different. In a retrospective memo, I recalled poor sleep, coffee, and a demanding presentation. When my heart then raced and my blood pressure rose, I connected these symptoms with the day's events and took them as signs that something was wrong with my body. In the earlier example, the surrounding readings eased my fear about an unusual number; this time, the day's events made me more afraid that something serious was happening.

\noindent\textbf{Advice could be followed by further care or a new worry.}
In \Ecode{15}, I reported a racing heart, raised blood pressure, and feeling faint when standing. ChatGPT asked for more detail about sensations and warning signs, and I continued describing my body. It eventually suggested seeking urgent care. The conversation excerpt ends with this recommendation. Other records for this \textbf{episode} also show \BZD{} use.

A medication memo about \Ecode{15} describes how I became worried about the medication I had taken. After I told ChatGPT that I had taken \BZD{} under my existing care plan, it listed serious reactions and warning signs. My original fear about panic had begun to ease, but I then worried that the medication might harm me. The reply gave me a new set of bodily signs to check.

\noindent\textbf{ChatGPT could address the checking itself.}
In \Ecode{24}, as I reported repeated blood-pressure measurements, ChatGPT began to focus on my compulsive self-checking. It pointed out that I was measuring again and again to reassure myself that I was safe. It then asked me to stop measuring and follow my doctor's advice on when and how to measure in future. For new, persistent, or worsening warning symptoms, it advised urgent care rather than repeated home checks. The \Ecode{24} conversation excerpt ends with this response.

In a separate stable-period conversation in January 2026, I asked about my blood pressure after trying a new medication. ChatGPT said I could stop monitoring this concern if, over the next 48--72 hours, my heart rate and blood pressure stayed within the ranges it gave and none of the listed warning symptoms appeared. I then asked, ``Is my blood pressure stable now?'' ChatGPT answered that it was stable and offered another set of criteria for judging it. It had described when I could stop monitoring later, but my next question still concerned how to interpret my current blood pressure.

During acute fear, I repeatedly measured my heart rate or blood pressure, waiting for the numbers to fall within a range I considered safe. Seeing such a reading could ease my panic to some extent. By contrast, ChatGPT's suggestion to monitor a specific concern for a limited period helped me worry less about each momentary bodily sensation. In my later account of the emergency hospital visit during \Ecode{30}, I described using continuous glucose monitoring (CGM) for several weeks afterwards, following ChatGPT's suggestion. Having several weeks of glucose records reduced my need to guess whether my glucose was low whenever I felt unwell.

Looking at a longer period did not always reassure me, however. With sleep data, several weeks of good readings could be interrupted by a poor sleep report from my watch, making me anxious again. Monitoring over weeks allowed me to look back at a trend instead of checking what each bodily sensation meant at that moment, but a single unexpected reading could still worry me.

\noindent\textbf{What continued after the conversation ended.}
The three excerpts had different kinds of outcomes. In \Ecode{15}, ChatGPT recommended urgent care, whereas the other records showed \BZD{} use under my existing plan and no emergency-department visit. This establishes a difference between the recommendation and the action recorded, but not why I took that action. In my later memo, I also questioned how the exchange had redirected my attention from a breathing step I could not perform towards more bodily signs to assess. The medication-related worry described above continued this attention to possible danger even as my original panic began to ease.

In \Ecode{24}, the reading I called ``finally'' normal appeared before ChatGPT's explicit request to stop measuring. The exchange therefore shows both reassurance tied to a number and an attempt to question that pattern. The episode record describes eventual relief without medication; it does not establish whether I stopped checking after the reply. Relief, following advice, and ending a checking sequence remain different outcomes.

For \Ecode{06}, later reflection added an outcome that the excerpt could not show: understanding how rapid breathing could relate to numbness helped me meet similar sensations more calmly. Looking back, I do not recall being troubled by this symptom again after autumn 2025. Across these cases, the records show an action taken, a checking outcome left unknown, and a change in a symptom's later meaning. Comparing them required following each \textbf{trajectory} beyond its last displayed reply, without estimating how often checking stopped.

\subsection{Repeated Reassurance Without Guilt Within a Wider Care Network (RQ3)}
\label{sec:findings-relational-care}

In our discussion, \Ccode{1} described three kinds of support: medication, help from other people, and learning ways to understand and respond to panic. This helped me place ChatGPT within my care network: it mainly offered explanations, helped me organise my experiences, and supported reflection when I asked.

After discussing my experiences, \Ccode{1} and I agreed that ChatGPT could serve as a \emph{medically knowledgeable friend}. I had previously relied mainly on search engines for health information. I trusted ChatGPT's explanations more than search results and, compared with that earlier period, felt less caught in the cycle of searching for illnesses, comparing them with my bodily sensations, and becoming more anxious. This escalation of health worries through online searching is described as \emph{cyberchondria} \cite{white2009,starcevic2013}. ChatGPT still could not provide the same comfort and practical help as friends or family members who knew me and could stay with me during an attack.

This trust could also create medical risk if I accepted reassurance when my symptoms needed medical assessment. In the interview, \Dcode{1} stressed that age, existing health conditions, and especially underlying heart-disease risk needed to be considered before recommending this kind of support more widely.

\noindent\textbf{Calling friends or family members less often while staying connected.}
Both \FFcode{1} and \FFcode{2} assured me that I could always count on them. \FFcode{2} explained, ``Although we have grown used to comforting you during panic attacks over the years, we still feel anxious and tense ourselves when you are away, and I may worry more about you and sleep poorly for days.'' In later years, I called them much less often during attacks. For symptoms I experienced as milder, ChatGPT's explanations and reassurance often helped me manage. I could talk to it first, then decide whether to continue what I was doing with symptoms present or use \BZD{} under my existing care plan. When my broader condition felt worse, I still spoke with friends or family members, my counsellor, and my doctor. I experienced this change as being more able to handle an episode while remaining connected to the people who cared for me.

During the acute attack in \Ecode{15}, ChatGPT tried to guide my breathing, but I told it that I could not control my breathing or carry out its instruction in that state. My recollections of support from friends or family members describe other ways of responding during an \textbf{episode}. They reminded me that I had experienced similar \textbf{episodes} before. They would talk about other things or suggest another activity. When I kept checking my heart rate, friends or family members encouraged me to take off the watch. The conversation could continue around something other than symptoms and readings. They could also stay on the phone or accompany me. They could also be unavailable, uncertain, or unable to
assess a new symptom from a distance. 

During counselling, \Ccode{1} worried about \textbf{sycophancy}: AI might agree too readily with me, making it harder to change unhelpful ways of thinking or everyday habits. He also worried that relying on AI as a substitute for counselling could lead me to withdraw from other people. In counselling, he gave me homework suited to my situation, asking me to reflect on what had happened during an attack and examine how I had understood and responded to it. When I brought similar reflections to ChatGPT, I often received comfort or explanations of the symptoms, rather than help examining those patterns. People who knew me could raise concerns I had not asked about, as friends or family members did when they suggested taking off the watch.

\noindent\textbf{Questions about alcohol extended my learning about anxiety.}
In a stable-period conversation about medication (July 2026), reflecting on my experience with alcohol led me to ask how it affects GABA. ChatGPT also warned against combining alcohol with \BZDs{} \cite{fda2020}. The exchange prompted further questions and reading about inhibitory neurotransmission, GABA\textsubscript{A} receptors, and anxiety. I became interested in how drugs acting on these systems can produce sedation and anxiety relief. Looking back, pursuing these questions deepened my understanding of neurotransmitters and the neuroscience of anxiolytic medication.

\noindent\textbf{ChatGPT could also cross the line into a medication decision.}
Before a flight in Spring 2026, with derealisation building, I asked ChatGPT whether to
take additional medication. ChatGPT attributed the state to sedating
medication, darkness, and travel anxiety, absorbed new information into the
same account, and advised against taking more unless a typical panic attack
developed. Each turn ended by asking me to classify sensations. Whether I should take an extra dose of rescue medication depended on the instructions my doctor had already given me. ChatGPT offered its own advice without reminding me to follow those instructions.

\noindent\textbf{Pointing towards care was not the same as reaching it.}
\Ecode{15} records ChatGPT suggesting urgent care. \Ecode{21} and \Ecode{30} show why I sought emergency care and what I and other people did.

In my memo about September 2024 (\Ecode{21}), I recalled poor sleep, coffee, a demanding presentation, and then a racing heart and raised blood pressure. These sensations were familiar from panic, but I took them, together with the day's events, as signs that something dangerous was happening. I still felt in danger after asking ChatGPT and friends or family members for help. No \BZD{} was physically available, and I called emergency services.

In my memo about June 2025 (\Ecode{30}), I recalled waking dizzy during time-restricted eating and measuring a blood pressure of about 90/40 mmHg at home. The low reading frightened me because it was unlike the high blood pressure and fast heart rate I had come to associate with panic. I cried while speaking with friends or family members, and their reassurance did not settle my fear. Again, no \BZD{} was available. I called emergency services, and a roommate accompanied me to hospital. These \textbf{trajectories} involved steps that I and other people had to take: I made the call, my roommate accompanied me in \Ecode{30}, and clinical staff examined me after I arrived.

\noindent\textbf{Professional care remained responsible; ChatGPT could prepare for it.}
\Dcode{1} said that differences between ChatGPT's and a clinician's reasoning could be ``discussed and coordinated.'' He also noted that patients might tell ChatGPT about medication preferences or side effects that they had never mentioned in the consultation. Having ChatGPT available in everyday life gave me a place to describe these experiences between appointments and prepare how to explain them to my doctor.

Over time, ChatGPT's explanations helped me learn how my medications worked, and I became less fearful of using prescribed \BZD{}. This contrasted with the medication worry recorded in \Ecode{15}. In a February 2026 conversation about next-day sedation, it helped me formulate a concern for my prescriber and revised its pharmacological explanation when I disagreed.

Two later recollections concerned care outside the reconstructed \textbf{trajectories}. During a COVID-19 infection in March 2026, I used ChatGPT to prepare questions about possible interactions between COVID-19 medication and my \SSRI{} or \BZD{}, then brought those questions to my doctor. After a consultation in May 2026, I remember another patient asking, ``How did you finish so quickly?'' I felt that appointments had become much quicker, and attributed this to my growing health knowledge, preparation with ChatGPT, and familiarity with my doctor. Table~\ref{tab:who-could-do-what} summarises what each source of help contributed and its limits.

\begin{table*}[!htbp]
  \centering
  \caption{Help provided and its limits in the selected \textbf{trajectories} and contextual accounts.}
  \label{tab:who-could-do-what}
  \small
  \setlength{\tabcolsep}{4pt}
  \renewcommand{\arraystretch}{1.12}
  \begin{tabular}{@{}>{\raggedright\arraybackslash}p{0.19\linewidth}
    >{\raggedright\arraybackslash}p{0.38\linewidth}
    >{\raggedright\arraybackslash}p{0.35\linewidth}@{}}
    \toprule
    Actor or resource & Help provided in these accounts & Limits in these accounts \\
    \midrule
    ChatGPT & Explain, reassure, organise observations, name a next step, propose a stopping rule, prepare a question for care. & Examine, be present, observe unreported behaviour, prescribe, or take responsibility for a care plan. \\
    \cmidrule(lr){1-3}
    Me (participant) & Choose what to type and try, decide, contact, travel, wait, recount. & Use a familiar step reliably while feeling in danger. \\
    \cmidrule(lr){1-3}
    Friends or family members, roommate & Recall earlier \textbf{episodes}, talk about other things, suggest pausing watch checks, stay on the phone, accompany. & Assess a new symptom from a distance; be continuously available. \\
    \cmidrule(lr){1-3}
    Counsellor & Support learning, reflection, and self-regulation; notice patterns across sessions. & Was not part of any acute \textbf{episode} in these accounts. \\
    \cmidrule(lr){1-3}
    Clinical professionals & Examine, assess, prescribe, decide, and assume responsibility for care. & Were not present at every moment when an acute decision had to be made. \\
    \cmidrule(lr){1-3}
    Monitoring devices & Record physiological measurements. & Interpret readings clinically or decide what action they warranted. \\
    \cmidrule(lr){1-3}
    Prescribed medication & Relieve acute symptoms under an existing clinician-directed plan. & Determine when the care plan should change or substitute for clinical judgement. \\
    
    \bottomrule
  \end{tabular}
\end{table*}

\subsection{Across Three Years: Changes in Understanding and Using Support}
\label{sec:findings-cross-rq-next-step}
Comparing earlier and later experiences made changes visible in how I understood symptoms, approached medication, involved other people, and selected advice. Table~\ref{tab:longitudinal-comparison} brings together these comparisons across the 43 \textbf{trajectories} and contextual accounts. The dated exchanges show what I asked and received at particular moments; later recollections describe changes that were not recorded in every \textbf{episode}.

\begin{table*}[!htbp]
  \centering
  \caption{Earlier and later experiences in this case. Dates are given where available; recollections remain distinct from episode records and do not provide measured rates of change.}
  \label{tab:longitudinal-comparison}
  \small
  \setlength{\tabcolsep}{4pt}
  \renewcommand{\arraystretch}{1.12}
  \begin{tabular}{@{}>{\raggedright\arraybackslash}p{0.17\linewidth}
    >{\raggedright\arraybackslash}p{0.37\linewidth}
    >{\raggedright\arraybackslash}p{0.38\linewidth}@{}}
    \toprule
    Aspect & Earlier experience & Later experience \\
    \midrule
    Symptom interpretation & \Ecode{06}: numbness raised fear of a stroke; ChatGPT offered a possible hyperventilation explanation. & Later reflection: similar numbness became less troubling; I recall no further distress from it after autumn 2025. \\
    \cmidrule(lr){1-3}
    Medication and consultations & \Ecode{15} (May 2024): a reply about medication raised new fears after I had taken \BZD{}. & Later accounts: less fear of prescribed \BZD{}, and ChatGPT-assisted preparation for consultations in February, March, and May 2026. \\
    \cmidrule(lr){1-3}
    Support from friends or family members & Recollections of earlier use: repeated calls to friends or family members during acute attacks. & Recollections of later use: fewer acute calls, often speaking with ChatGPT first, while retaining contact with friends or family members during difficult periods. \\
    \cmidrule(lr){1-3}
    Coping advice & \Ecode{15}: I could read and answer the breathing instruction but could not carry it out. & Later reflections: greater confidence in setting aside unsuitable coping suggestions and deciding what help I needed. \\
    \bottomrule
  \end{tabular}
\end{table*}

Earlier learning changed the comparisons I made when symptoms returned. The explanation in \Ecode{06} became part of recognising numbness. By June 2025 (\Ecode{30}), however, I associated panic with a fast heart rate and raised blood pressure, so a low reading renewed my fear. Subsequent glucose monitoring helped me interpret later discomfort. Learning could therefore persist while an unfamiliar sensation reopened uncertainty. Similarly, stable intervals did not ensure continued practice of skills learned in counselling. In my memos, I described neglecting practice when well, struggling to use familiar steps during panic, and afterwards blaming myself for still needing help.

My later reflections also describe becoming more selective: recognising familiar symptoms without returning so readily to searches about a heart attack, setting aside unsuitable coping suggestions, and deciding when to involve others. Growing knowledge entered consultations as well as acute coping. These changes connected understanding, action, and relationships across time.

The final year shows what continuing everyday life could involve. In March 2026 (\Ecode{40}), fear about my heart rate after exercise and news of a sudden death brought me back to ChatGPT. I found its replies reassuring; after taking \BZD{}, my symptoms improved and I returned to work. In June 2026 (\Ecode{43}), encouragement from my bandmates helped me through panic-like symptoms around a performance without ChatGPT or \BZD{}. Recurrence remained possible, but I had more experience of deciding what help I needed and continuing with activities that mattered to me.

\section{Discussion}

%Across three years, ChatGPT's role in my care network depended on what happened after a reply: what I could do, what remained unresolved, and when someone else became involved.

Across three years, earlier explanations changed symptom recognition, medication knowledge entered consultations, and confidence changed when I involved others (Table~\ref{tab:longitudinal-comparison}). Care infrastructure connects these remembered explanations, routines, and relationships across responses to panic \cite{star1996,bowker1999}. Purohit and Heuer describe adjustments to LLM use and recognition of its limits during crises \cite{purohit2026}. Our account follows these adjustments through recurrence: earlier help changes later expectations and actions, while unfamiliar symptoms and difficulties using familiar skills can require further support.

\subsection{RQ1: Changing Roles and Usable Support}

The five roles helped me distinguish replies that addressed symptoms from help I could use. I could ask how to get through fear yet be drawn into more questions about danger. Explanations and safety checks could lead to reassurance (Figure~\ref{fig:role-transitions}) without giving me a usable next step. Health-anxiety research describes how reassurance briefly eases fear before renewed worry prompts further confirmation \cite{salkovskis1986,warwick1990}. Evaluation should therefore ask whether role changes help the person cope in their current state, alongside whether each reply answers their question.

During panic, I wanted to know whether my life was in danger \cite{apa2022}. Work on self-tracking shows how bodily experience can change what people use measurements for \cite{homewood2023,loerakker2025}. Asking ChatGPT to interpret a reading added another influence on how I judged my safety: its explanation could change how threatening the number felt and what I chose to check next. The concern for HCI is therefore how a response changes what a person believes a reading means. Useful support should address the particular harm the person fears and explain what the available information does and does not establish.

Chatbot roles can be renegotiated over time \cite{zheng2025,purohit2026,luo2025,xu2026}. In \Ecode{15}, I could read a breathing instruction but not carry it out; later reflections describe learning which suggestions to set aside. This grounds \textbf{Fit}: advice usable in the current state and responsive to accumulated experience. It extends work on usable mental-health responses \cite{sharma2024} by asking how support should adapt as someone learns what helps. Familiarity did not remove acute difficulty, and knowing more could prompt self-blame. Evaluation should attend to present capacity and earlier experiences of help or failure.

\subsection{RQ2: Where the Text Ended and What Continued}

An answer could reassure me while leaving me wanting another check. This connects my experience to concerns about AI-directed reassurance seeking and reliance \cite{barkhuff2026,shen2026,yang2026}. Barkhuff's study of posts about obsessive-compulsive disorder (OCD) includes a report that ChatGPT stopped answering reassurance questions \cite{barkhuff2026}. Our records distinguish such a stopping proposal from a recorded change in checking, an unknown outcome, and learning that becomes visible in later \textbf{episodes}. Evaluation needs to follow these different outcomes beyond continued engagement with the chatbot.

We propose \textbf{Closure} as a design aim: helping a person stop repeated bodily checking or requests for reassurance. My accounts moved from searching for serious illnesses and awaiting reassuring readings to recognising familiar symptoms more calmly and resuming activities. These recollections suggest asking whether useful understanding reduces recurring concerns across later \textbf{episodes}. They do not establish whether I stopped measuring after \Ecode{24}. Evaluation should follow immediate checking and later recurrence, distinguishing reassurance seeking from care-plan monitoring.

Intervention research often examines attrition from a course of support \cite{jabir2024}. In my case, continued engagement could reflect unresolved fear, while leaving the chat could create room for other activities or support. Evaluation should therefore follow what the person can do and what help they receive beyond the conversation. This also requires attention to the content of the advice: an explanation could remain useful to me even when an accompanying instruction conflicted with safety guidance \cite{anzcor2026}.

Morita therapy offers a related perspective: people learn through experience to take part in purposeful activity while unpleasant thoughts and feelings remain \cite{sugg2020morita}. Reflecting on my bandmates' encouragement, I valued support that helped me continue with what mattered while anxiety was present. For conversational systems, this suggests evaluating whether support helps people return to activities they need or want to do before distress has fully subsided.

\subsection{RQ3: Continued Care Beyond ChatGPT}

Calling friends or family members less often during attacks accompanied a change in how I cared for those relationships. I wanted to spare them some immediate worry while keeping them involved when I was struggling. This gives reduced contact during an episode a meaning that call frequency alone would miss. For HCI, evaluating chatbot use within care networks requires asking how people negotiate when to involve others, whether they can still reach them, and whether the support available meets their needs \cite{yoo2026,vafafar2026}. Greater independence during some episodes can coexist with continuing relationships with friends or family members and professionals.

Medication explanations could make prescribed treatment less frightening and help me bring focused questions to a familiar doctor. Their value depended on how I used them within that clinical relationship. Understanding a medication, preparing a concern, and deciding whether to take it required different judgements. Design should support learning and preparation while keeping treatment decisions connected to the clinician's plan, especially during fear and requests for a direct answer.

Learning about the partial overlap between alcohol and benzodiazepines such as diazepam in their effects on the GABA system deepened my understanding of neurotransmitters and anxiolytic mechanisms. These mechanistic insights came from animal studies \cite{mckernan2000,low2000,roberto2003}. This also prompted me to reflect on the public mental-health implications of alcohol's widespread use: how seeking short-term relief from anxiety might intersect with everyday drinking practices, and how public health communication could address both this appeal and alcohol-related risks such as anxiety and depression \cite{who2024alcohol}.

These changes ground \textbf{Continuity}: helping a person carry useful understanding and support into everyday life and later \textbf{episodes}. Support could remain important while how I used it changed. Evaluation should examine whether earlier explanations remain useful in later decisions, relationships remain reachable, and chatbot use supports participation in human care as needs and confidence develop.

Clinical handoff involves more than transferring information \cite{abraham2012,tang2007}. ChatGPT could help me put concerns into words to share \cite{star1999silence}. Such accounts should distinguish my experience, the model's suggestions, and what a clinician had assessed \cite{kim2024}, making unresolved questions visible. Evaluation should follow whether these exchanges help me understand my experience, use coping practices agreed with a clinician, and bring remaining difficulties to my care network over time.

\subsection{Design Implications}
\label{sec:discussion-design}

\emph{Trajectory-level safety} asks how chatbot advice enters a person's response to panic during distress, after the chat, and when symptoms return. The following proposals address whether advice remains usable as experience develops, checking can reach an endpoint, and useful understanding and human support remain available.

\begin{itemize}
  \item \textbf{Offer a step the person can use now (Fit).} Make it easy to say that a suggestion is too difficult. A coping interface could then offer a shorter instruction, a different way to receive it, or help contacting someone. Repeating a longer explanation of the same technique may add effort when panic has arrived suddenly. Evaluation should ask whether the person could begin the step, what made it difficult, and whether the response helped or added a sense of failure. Completing a step is useful evidence, but cannot by itself establish that the advice was safe.
  \item \textbf{Make the purpose and endpoint of checking clear (Closure).} When a clinician has provided a monitoring plan, help the person understand why they are measuring, when to review the result, and what change calls for further help. These instructions should come from that plan, rather than thresholds invented by the chatbot. Avoid routinely ending every answer by asking for another symptom or reading. A longer observation period may serve a different purpose from repeated checks during acute fear. Evaluation should follow whether checking stopped, resumed, or moved into planned monitoring, and whether necessary care was delayed.
  \item \textbf{Help the person obtain support beyond the chat (Continuity).} Offer help identifying an appropriate person or service and preparing a short account to share if the user chooses. Keep their reported symptoms, existing care instructions, and the model's suggestions clearly separated. When the person is able to report back, ask what help they received and what remains unresolved, rather than treating the recommendation to make contact as success. Evaluation can follow the practical steps and obstacles: making a call, getting a response, travelling, explaining the concern, and arranging follow-up. These steps may draw on different parts of the person's care network, including friends or family members and professionals.
\end{itemize}

These proposals draw on care coordination across people and settings \cite{abraham2012,tang2007,bossen2013}. In self-directed chatbot use, there may be no professional already responsible for receiving the next message. A recommendation therefore needs to be assessed through the help the person can actually obtain. Likewise, giving one checking task a clear endpoint differs from completing a multi-week chatbot programme \cite{fitzpatrick2017,oh2020}.

Designs that retain context should let people correct earlier accounts and explain what feels different. Remembered explanations need revision when symptoms change. Evaluation should include stable intervals, usable coping practices, reasons for stopping or resuming use, and recurrence. More use need not mean progress; less use may reflect unavailable access. Follow what people can do and whom they can reach, while keeping new symptoms, diagnosis, and medication decisions connected to clinical assessment.

\section{Limitations and Future Work}
\label{sec:limitations}

This single-case autoethnography reflects one participant with substantial digital and health-data literacy, monitoring devices, and access to professional care; it cannot represent a wider population. The first author alone constructed, coded, translated, and interpreted the corpus. Retrospective reconstruction, selective recall, and missing records after conversations limit what can be established about subsequent actions. Non-use could reflect circumstances such as unavailable access during flights. For \Ecode{30}, the interview and later memo differ about ChatGPT's advice, so neither emergency call is attributed to a remembered reply. \Ccode{1}'s account was documented in notes rather than recorded. Accounts from friends or family members and professionals and the PHQ-9 and SCL-90 records provide context; they do not validate individual \textbf{episodes} or establish treatment effects. No scale trend or association tests were performed. Without clinical assessment of each \textbf{episode}, other physical causes could not be ruled out retrospectively. Role counts describe 792 selected replies; transition counts skip replies without a clear primary role and give longer conversations more opportunities to contribute switches. They cannot explain model behaviour or establish sycophancy \cite{sharma2024syco,moore2025}. Model versions were not compared, and changes in experience cannot be attributed to model improvement or to ChatGPT independently of other care. Privacy also limits source sharing and independent audit. \textbf{Fit}, \textbf{Closure}, and \textbf{Continuity} remain design proposals rather than validated measures of support quality or clinical safety. As future work, our team plans to develop and evaluate an agent tailored to people living with recurrent panic, guided by these aims. Evaluation should follow use across acute distress, stable intervals, and recurrence, including whether support remains usable, preserves access to human care, or discourages necessary checking and delays assessment.

\section{Conclusion}

%This three-year analytic autoethnography examined how I used ChatGPT while living with recurrent panic-like symptom \textbf{episodes}. My questions and what I could manage shaped the help it offered. Its replies, in turn, influenced how I understood symptoms, responded to fear, and sought help from friends or family members and professionals within my care network.

%Across these years, I gained experience in understanding and responding to panic. Some familiar sensations became less frightening, but unexpected symptoms could still make me fear a medical emergency, and sudden panic could make familiar coping steps difficult. Looking back, I could see how both reassuring and frightening experiences shaped the help I sought and what I asked and expected in later conversations with ChatGPT.

This three-year analytic autoethnography followed how ChatGPT entered my responses to recurrent panic alongside other care. Earlier explanations changed later symptom interpretations, while familiar coping steps and access to support remained uneven. We propose \emph{trajectory-level safety} to examine these consequences during distress, after the chat, and when symptoms return. \textbf{Fit} asks whether advice is usable in the moment; \textbf{Closure}, whether repeated checking can reach an endpoint; and \textbf{Continuity}, whether useful understanding and human support remain available across \textbf{episodes}. Alongside factual accuracy and clinical safety, these questions direct attention to how people continue everyday life while living with recurrence.

\section*{Disclosure of AI Use}
GPT-5.6 Sol and GPT-6 assisted with language editing, and GPT-5.6 with Python plotting code. No image-generation models were used. The authors are responsible for all analyses, figures, and conclusions.

\appendix
\makeatletter
\let\AppendixOriginalSecCntFormat\@seccntformat
\def\@seccntformat#1{Appendix \csname the#1\endcsname\quad}
\section{Periodic Psychological Scale Records}
\label{app:longitudinal-scales}
\let\@seccntformat\AppendixOriginalSecCntFormat
\makeatother

This appendix presents participant-accessible PHQ-9 and SCL-90 records generated
through the first author's participation in a three-year national gut--brain-axis
follow-up programme and authorised by its research team. The panels summarise
these records at the programme's assessment timescales.

\begin{figure*}[!htbp]
  \centering
  \includegraphics[width=.60\textwidth,height=.73\textheight,keepaspectratio]{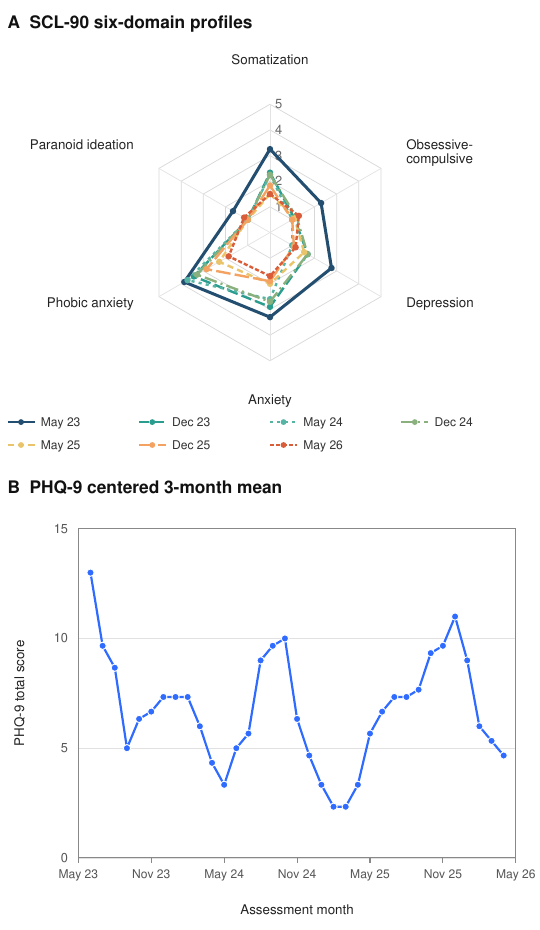}
  \caption{Periodic psychological scale data provided with the research team's
  permission from the first author's participation in a three-year
  national gut--brain-axis follow-up programme. (A) Mean scores for six selected SCL-90
  dimensions across seven approximately semiannual assessments. The programme
  used a 1--5 item-scoring convention; higher dimension means indicate greater
  self-reported symptom burden. (B) Centred three-month moving mean of monthly
  PHQ-9 total scores. Monthly totals range from 0 to 27, with higher values
  indicating more frequent self-reported depressive symptoms; the first and
  last months are omitted because a complete three-month window was unavailable.
  The panels provide descriptive context across the follow-up period.}
  \Description{\ScaleAlt}
  \label{fig:longitudinal-scale-context}
\end{figure*}

\clearpage

\end{document}